\documentclass[12pt]{article} 
\usepackage[sectionbib]{natbib}
\usepackage{array,epsfig,fancyheadings,rotating}
\usepackage[]{hyperref}  
\usepackage{sectsty, secdot}
\sectionfont{\fontsize{12}{14pt plus.8pt minus .6pt}\selectfont}
\renewcommand{\theequation}{\thesection\arabic{equation}}
\subsectionfont{\fontsize{12}{14pt plus.8pt minus .6pt}\selectfont}

\usepackage{amsmath}
\usepackage{amssymb}
\usepackage{amsfonts}
\usepackage{multirow}
\usepackage{amsthm}
\usepackage{subcaption}
\usepackage{xcolor}

\newtheorem{theorem}{Theorem}

\theoremstyle{definition}

\def\bSig\mathbf{\Sigma}

\DeclareMathOperator\bE{\mathbb E} 
\DeclareMathOperator\bV{\mathbb V} 
\newcommand{\bcx}{{\bf X}}

\newcommand{\bci}{{\bf I}}

\newcommand{\bcs}{{\bf S}}

\newcommand{\bfbeta}{\mbox{\boldmath $\beta$}}
\newcommand{\bfalpha}{\mbox{\boldmath $\alpha$}}

\newcommand{\balpha}{\mbox{\boldmath $\alpha$}}

\newcommand{\bfgamma}{\mbox{\boldmath $\gamma$}}

\newcommand{\indep}{\perp \!\!\! \perp}

\newtheorem{Remark}{\sc Remark}

\begin{document}


\renewcommand{\baselinestretch}{2}

\markright{ \hbox{\footnotesize\rm Statistica Sinica
}\hfill\\[-13pt]
\hbox{\footnotesize\rm
}\hfill }

\markboth{\hfill{\footnotesize\rm Shuxi Zeng and others} \hfill}
{\hfill {\footnotesize\rm Propensity score weighting for survival outcomes} \hfill}

\renewcommand{\thefootnote}{}
$\ $\par


\fontsize{12}{14pt plus.8pt minus .6pt}\selectfont \vspace{0.8pc}
\centerline{\large\bf Propensity Score Weighting Analysis of Survival }
\vspace{2pt} 
\centerline{\large\bf Outcomes Using Pseudo-observations }
\vspace{.4cm} 
\centerline{Shuxi Zeng$^{1}$, Fan Li $^{1}$, Liangyuan Hu $^{2}$, Fan Li $^{3,\ast}$} 
\vspace{.4cm} 
\centerline{\it $^{1}$Duke University, $^{2}$Rutgers School of Public Health}
\centerline{\it and 
$^{3}$Yale School of Public Health}
\centerline{ $^{\ast}$fan.f.li@yale.edu}
\vspace{.55cm} \fontsize{9}{11.5pt plus.8pt minus.6pt}\selectfont


\begin{quotation}
\noindent {\it Abstract:}
Survival outcomes are common in comparative effectiveness studies and require unique handling because they are usually incompletely observed due to right-censoring. A ``once for all'' approach for causal inference with survival outcomes constructs pseudo-observations and allows standard methods such as propensity score weighting to proceed as if the outcomes are completely observed. For a general class of model-free causal estimands with survival outcomes on user-specified target populations, we develop corresponding propensity score weighting estimators based on the pseudo-observations and establish their asymptotic properties. In particular, utilizing the functional delta-method and the von Mises expansion, we derive a new closed-form variance of the weighting estimator that takes into account the uncertainty due to both pseudo-observation calculation and propensity score estimation. This allows valid and computationally efficient inference without resampling. We also prove the optimal efficiency property of the overlap weights within the class of balancing weights for survival outcomes. The proposed methods are applicable to both binary and multiple treatments. Extensive simulations are conducted to explore the operating characteristics of the proposed method versus other commonly used alternatives. We apply the proposed method to compare the causal effects of three popular treatment approaches for prostate cancer patients. 
\vspace{9pt}
\par
\end{quotation}\par

\def\thefigure{\arabic{figure}}
\def\thetable{\arabic{table}}

\renewcommand{\theequation}{\thesection.\arabic{equation}}

\fontsize{12}{14pt plus.8pt minus .6pt}\selectfont

\section{Introduction}\label{s:intro}

Survival or time-to-event outcomes are common in comparative effectiveness research and require unique handling because they are usually incompletely observed due to right-censoring. In observational studies, a popular approach to draw causal inference with survival outcomes is to combine standard survival estimators with propensity score methods \citep{Rosenbaum83}. For example, one can construct the Kaplan-Meier estimator on an inverse probability weighted sample to adjust for measured confounding \citep{robins2000correcting_IPC,hubbard2000nonparametric}. Another common approach combines the Cox model with inverse probability weighting (IPW) to estimate the causal hazard ratio \citep{austin2017performance} or the counterfactual survival curves \citep{cole2004adjusted}. 
Coupling causal inference with the Cox model introduces two limitations. First, the Cox model assumes proportional hazards in the target population, violation to which leads to biased causal estimates. Second, the target estimand is usually the causal hazard ratio, whose interpretation can be opaque due to the built-in selection bias \citep{hernan2010hazards}. In contrast, estimands based on survival probability or restricted mean survival time are free of model assumptions and have a natural causal interpretation \citep{mao2018propensitysurvival}. 

To analyze observational studies with survival outcomes, an attractive alternative approach is to combine causal inference methods with \emph{pseudo-observations} \citep{andersen2003generalisedpseudo}. Each pseudo-observation is constructed based on a jackknife statistic and is interpreted as the individual contribution to the target estimate from a complete sample without censoring. The pseudo-observations approach addresses censoring in a ``once for all'' manner and allows standard methods to proceed as if the outcomes are completely observed \citep{andersen2004regressionpseudo}. To this end, one can perform direct confounding adjustment using outcome regression with pseudo-observations and derive casual estimators with the g-formula \citep{robins1986new}. Another approach is to combine propensity score weighting with pseudo-observations. \citet{andersen2017causalpseudo} considered an IPW estimator to estimate the causal risk difference and difference in restricted mean survival time. Their approach was further extended to doubly robust estimation with survival and recurrent event outcomes \citep{wang2018doublerobustsurvival,su2020causal}.

Despite its simplicity and versatility, several open questions in propensity score weighting with pseudo-observations remain to be addressed. First, pseudo-observations require computing a jackknife statistic for each unit, which poses computational challenges to resampling-based variance estimation under propensity score weighting \citep{andersen2017causalpseudo}. On the other hand, failure to account for the uncertainty in estimating the propensity scores and jackknifing can lead to inaccurate and often conservative variance estimates. Second, the IPW estimator with pseudo-observations corresponds to a target population that is represented by the study sample, but the interpretation of such a population is often questionable in the case of a convenience sample \citep{li2019addressingow}. Moreover, the inverse probability weights are prone to lack of covariate overlap and will engender causal estimates with excessive variance, even when combined with outcome regression \citep{mao2019propensity}. \citet{li2018balancing} proposed a general class of balancing weights (which includes the IPW as a special case) to define target estimands 
on user-specified target populations. In particular, the overlap weights emphasize a target population with the most covariate overlap and best clinical equipoise, and were theoretically shown to provide the most efficient causal contrasts. However, the theory of overlap weights so far has focused on non-censored outcomes, and its optimal variance property is unclear with survival outcomes. Third, many comparative effectiveness studies involve multiple treatments, which can exacerbate the consequence of lack of overlap when IPW is considered \citep{yang2016multi}. 
While the overlap weights \citep{li2019propensity} offered a promising solution to improve the bias and efficiency over IPW with non-censored outcomes, extensions to censored survival outcomes remain limited, with one exception of \citet{cheng2022addressing} for binary treatments. 

In this paper, we address the above questions. We consider a general multiple treatment setup and extend the balancing weights in \citet{li2018balancing} and \citet{li2019propensity} to analyze survival outcomes in observational studies based on the pseudo-observations. We develop new asymptotic variance expressions for causal effect estimators that account for the variability due to both estimating propensity scores and constructing pseudo-observations. Different from existing variance expressions developed for propensity score weighting estimators \citep{lunceford2004stratification,mao2018propensitysurvival}, our new asymptotic variance expression is developed based on the functional delta-method and the von Mises expansion \citep{graw2009pseudo, jacobsen2016pseudo_large_sample, overgaard2017asymptotic}, which are uniquely required in this context as the pseudo-observations are themselves estimated via jackknifing. Such asymptotic results also enable valid and computationally efficient inference without re-sampling. Based on the new asymptotic variance expression, we further prove that overlap weights lead to the most efficient survival causal estimators, expanding the theoretical underpinnings of overlap weights to causal survival analysis. We carry out simulations to evaluate and compare a range of commonly used weighting estimators. Finally, we apply the proposed method to estimate the causal effects of three treatments on mortality among patients with high-risk localized prostate cancer from the National Cancer Database.

\section{Propensity score weighting with survival outcomes}\label{s:psw}
\subsection{Time-to-event outcomes, causal estimands and assumptions}\label{s:setup}

We consider a sample of $N$ units drawn from a population. Let $Z_{i}\in\mathcal{J}=\{1,2,\cdots,J\}, J\geq 2$ denote the assigned treatment. Each unit has a set of potential outcomes $\{T_{i}(j),j\in\mathcal{J}\}$, measuring the counterfactual survival time mapped to each treatment. We similarly define $\{C_{i}(j),j\in\mathcal{J}\}$ as a set of potential censoring times. Under the Stable Unit Treatment Value Assumption (SUTVA), $T_i=\sum_{j\in\mathcal{J}}\textbf{1}\{Z_{i}=j\}T_i(j)$ and $C_i=\sum_{j\in\mathcal{J}}\textbf{1}\{Z_{i}=j\}C_i(j)$. Due to right-censoring, we might only observe the lower bound of the survival time for some units. We write the observed failure time, $\widetilde{T}_{i}= T_{i}\wedge C_{i}$, the censoring indicator, $\Delta_{i}=\textbf{1}\{T_{i}\leq C_{i}\}$, and the $p$-dimensional time-invariant pre-treatment covariates, $\bcx_{i}=(X_{i1},\ldots,X_{ip})'\in \mathcal{X}$. In summary, we observe the tuple  $\mathcal{O}_i=(Z_{i},\bcx_{i},\widetilde{T}_{i},\Delta_{i})$ for each unit. 
We define the generalized propensity score, $e_{j}(\bcx_{i})=\Pr(Z_{i}=j|\bcx_{i})$, as the probability of receiving treatment $j$ given baseline covariates \citep{imbens2000role}. Our results are presented for general, finite $J\geq 2$. 

The causal estimands of interest are based on two typical transformations of the potential survival times: (i) the at-risk function, $\nu_{1}(T_{i}(j);t)=\textbf{1}\{T_{i}(j)\geq t\}$, and (ii) the truncation function, $\nu_{2}(T_{i}(j);t)=T_{i}(j)\wedge t$, where $t$ is a given time point of interest. The identity function is implied by $\nu_{2}(T_{i}(j);\infty)=T_{i}(j)$. To simplify the discussion, hereafter we use $k\in\{1,2\}$ to index the choice of the transformation function $v$. We further define $m_{j}^{k}(\bcx;t)=\bE\{\nu_{k}(T_{i}(j);t)|\bcx\}$ as the conditional expectation of the transformed potential survival outcome, and the pairwise conditional causal effect at time $t$ as $\tau_{j,j'}^{k}(\bcx;t)=m_{j}^{k}(\bcx;t)-m_{j'}^{k}(\bcx;t)$ for $j\neq j' \in\mathcal{J}$. We are interested in the conditional causal effect averaged over a \emph{target population}. We assume the study sample is drawn from the population with covariate density $f(\bcx)$ (with respect to a measure  $\mu(\cdot)$), and represent the target population by density $g(\bcx)$. The function $h(\bcx)\propto g(\bcx)/f(\bcx)$ is a tilting function, which re-weights the observed sample to represent the target population. The \emph{pairwise average causal effect} at time $t$ on the target population is defined as 
\begin{eqnarray}
\label{eq:Arm-average}
\tau_{j,j'}^{k,h}(t)=\frac{\int_{\mathcal{X}}\tau_{j,j'}^{k}(\bcx;t)f(\bcx)h(\bcx)\mu(d\bcx)}{\int_{\mathcal{X}}f(\bcx)h(\bcx)\mu(d\bcx)},~~\forall~~j\neq j'\in\mathcal{J}.
\end{eqnarray}

The class of estimands \eqref{eq:Arm-average} is transitive in the sense that $\tau_{j,j'}^{k,h}(t)=\tau_{j,j''}^{k,h}(t)+\tau_{j'',j'}^{k,h}(t)$. Different choices of function $\nu_{k}$ lead to estimands on different scales. When $k=1$, we refer to estimand \eqref{eq:Arm-average} as the survival probability causal effect (SPCE). This estimand represents the causal risk difference and contrasts the potential survival probabilities at time $t$ among the target population. When $k=2$, estimand \eqref{eq:Arm-average} is referred to as the restricted average causal effect (RACE), which compares the mean potential survival times restricted by $t$. When $t=\infty$, this estimand becomes the average survival causal effect (ASCE) comparing the unrestricted mean potential survival times. However, because the observed data do not contain information beyond the maximum follow-up time $t_{\text{max}}$, one can at most identify $\tau_{j,j'}^{k=2,h}(t_{\text{max}})$. With sufficiently long follow-up time as in our data example, a practical solution is to estimate $\tau_{j,j'}^{k=2,h}(t_{\text{max}})$ (RACE at time $t_{\text{max}}$) as an approximation to ASCE. In this sense, the following inferential details for RACE still applies to ASCE. We will also examine the accuracy of this strategy to estimate ASCE in our simulations. Finally, when $J=2$, estimands \eqref{eq:Arm-average} reduce to those in \citet{mao2018propensitysurvival} for binary treatments.

To identify estimands \eqref{eq:Arm-average}, we maintain the following assumptions. For each $j\in \mathcal{J}$, we assume (A1) weak unconfoundedness: $T_{i}(j) \indep \textbf{1}\{Z_{i}=j\} |\bcx_{i}$; (A2) overlap: $0<e_{j}(\bcx)<1$ for any $\bcx\in\mathcal{X}$; and (A3) completely independent censoring: $\{T_{i}(j),Z_{i},\bcx_{i}\}\indep C_{i}(j)$. Assumption (A1) and (A2) are the usual no unmeasured confounding and positivity conditions typically invoked for multiple treatments \citep{imbens2000role,yang2016multi}, and allow us to identify $\tau_{j}^{k,h}(t)$ in the absence of censoring. Assumption (A3) assumes that censoring is independent of all remaining variables, and is introduced for now as a convenient technical device to establish our main results. (A3) often holds, for example, when the failure times are only subject to administrative right censoring. We will relax this assumption in Section \ref{s:theoretical} and \ref{s:simulations} to enable identification under a weaker condition, which assumes (A4) covariate dependent censoring: $T_{i}(j)\indep C_{i}(j)|\bcx_{i},Z_i=j$.

\subsection{Balancing weights with pseudo-observations}\label{sec:PO}
We now introduce balancing weights to estimate the causal estimands \eqref{eq:Arm-average}. Write $f_j(\bcx)=f(\bcx|Z=j)$ as the conditional density of covariates among treatment group $j$ over $\mathcal{X}$. It is immediate that $f_j(\bcx)\propto f(\bcx)e_j(\bcx)$. For any pre-specified tilting function $h(\bcx)$, we weight the group-specific density to the target population density using the following balancing weights, up to a proportionality constant:
\begin{align}
w^h_j(\bcx)\propto\frac{g(\bcx)}{f_j(\bcx)}\propto\frac{f(\bcx)h(\bcx)}{f(\bcx)e_j(\bcx)}=\frac{h(\bcx)}{e_j(\bcx)},~~\forall~~j\in \mathcal{J}.
\end{align}
The set of weights $\{w_j^h(\bcx):j\in\mathcal{J}\}$ balance the weighted distributions of pre-treatment covariates towards the corresponding target population distribution, i.e., $f_j(\bcx)w^h_j(\bcx)\propto g(\bcx)$, for all $j\in\mathcal{J}$.

To apply the balancing weights to survival outcomes subject to right-censoring, we first construct the pseudo-observations \citep{andersen2003generalisedpseudo}. For a given time $t$, we generically define $\theta^{k}(t)=\bE\{\nu_{k}({T}_{i};t)\}$ as a population parameter. The pseudo-observation for each unit is written as $\widehat{\theta}_{i}^{k}(t)=N\widehat{\theta}^{k}(t)-(N-1)\widehat{\theta}_{-i}^{k}(t)$, where $\widehat{\theta}^{k}(t)$ is the consistent estimator of $\theta^{k}(t)$, and $\widehat{\theta}_{-i}^{k}(t)$ is the corresponding estimator with unit $i$ left out. For transformation $\nu_{k}$ $(k=1,2)$, we consider the Kaplan–Meier estimator to construct $\theta^{k}(t)$, given by
$\widehat{S}(t)=\prod_{\widetilde{T}_i\leq t}\left\{1-\frac{dN(\widetilde{T}_i)}{Y(\widetilde{T}_i)}\right\},$
where $N(t)=\sum_{i=1}^N\textbf{1}\{\widetilde{T}_i\leq t,\Delta_i=1\}$ is the counting process for the event of interest, and $Y(t)=\sum_{i=1}^N\textbf{1}\{\widetilde{T}_i\geq t\}$ is the at-risk process. When the interest lies in the survival functions ($k=1$), the $i$th pseudo-observation is estimated by
$\widehat{\theta}_{i}^{1}(t)=N \widehat{S}(t)-(N-1)\widehat{S}_{-i}(t).$
When the interest lies in the restricted mean survival times ($k=2$), the $i$th pseudo-observation is estimated by
$\widehat{\theta}_{i}^{2}(t)=N\int_0^t \widehat{S}(u)du-(N-1)\int_0^t \widehat{S}_{-i}(u)du=\int_0^t \widehat{\theta}_{i}^{1}(u)du.$
The pseudo-observation is a leave-one-out jackknife approach to address right-censoring and provides a straightforward unbiased estimator of the functional of uncensored data under the independent censoring assumption (A3). From \citet{graw2009pseudo} and \citet{andersen2017causalpseudo} and under the unconfoundedness assumption (A1), one can show that $\bE\{\widehat{\theta}_{i}^{k}(t)|\bcx_{i},Z_{i}=j\}\approx\bE\{\nu_{k}(T_{i};t)|\bcx_{i},Z_{i}=j\}=\bE\{\nu_k(T_i(j);t)|\bcx_i\}$,
based on which the g-formula can be used to estimate the pairwise average causal effect on the overall population ($h(\bcx)=1$). For the class of estimands \eqref{eq:Arm-average}, we further propose the following nonparametric H\'{a}jek-type estimator:
\begin{eqnarray}
\label{eq:PSW_Pseudo}
\widehat{\tau}_{j,j'}^{k,h}(t)=\frac{\sum_{i=1}^{N}\textbf{1}\{Z_{i}=j\}\widehat{\theta}_{i}^{k}(t)w_{j}^{h}(\bcx_{i})}
{\sum_{i=1}^{N}\textbf{1}\{Z_{i}=j\}w_{j}^{h}(\bcx_{i})}-
\frac{\sum_{i=1}^{N}\textbf{1}\{Z_{i}=j'\}\widehat{\theta}_{i}^{k}(t)w_{j'}^{h}(\bcx_{i})}
{\sum_{i=1}^{N}\textbf{1}\{Z_{i}=j'\}w_{j'}^{h}(\bcx_{i})}
\end{eqnarray}

Estimator \eqref{eq:PSW_Pseudo} compares the weighted average pseudo-observations in each treatment group. 
First, without censoring, the $i$th pseudo-observation is simply the transformation of the observed outcome $\nu_k(T_i;t)$, and \eqref{eq:PSW_Pseudo} is identical to the estimator in \citet{li2019propensity} for complete outcomes. Second, a number of weighting schemes proposed for non-censored outcomes are applicable to \eqref{eq:PSW_Pseudo}. For example, the IPW estimator considers $h(\bcx)=1$ and $w_j^h(\bcx)=1/e_j(\bcx)$, corresponding to a target population of the combination of all treatment groups represented by the study sample. In this case, when only $J=2$ treatments are present, estimator \eqref{eq:PSW_Pseudo} reduces to the IPW estimator in \citet{andersen2017causalpseudo}. When the target population is the group receiving treatment $l$ (similar to the average treatment effects for the treated estimand in binary treatments), the corresponding $h(\bcx)=e_l(\bcx)$ and the balancing weight is $w_j^h(\bcx)=e_l(\bcx)/e_j(\bcx)$. The overlap weights (OW) specify $h(\bcx)=\left\{\sum_{l\in\mathcal{J}}e_{l}^{-1}(\bcx)\right\}^{-1}$ and $w_{j}^{h}(\bcx)=e_{j}^{-1}(\bcx)\left\{\sum_{l\in\mathcal{J}}e_{l}^{-1}(\bcx)\right\}^{-1}$, and correspond to the target population as an intersection of all treatment groups with optimal covariate overlap \citep{li2019propensity}. This overlap population mimics that enrolled in a randomized trial and emphasizes units whose treatment decisions are most ambiguous. When different groups have good covariate overlap, OW and IPW correspond to almost identical target population and estimands. The difference in target population and estimands between OW and IPW emerges with increasing regions of poor overlap. Specifically, as $e_j(\bcx)$ approaches zero, $w_j^h(\bcx)$ under IPW increases to infinity, whereas $w_j^h(\bcx)$ under OW approaches to zero. Due to such intrinsic differences in construction of the weights, OW is expected to improve efficiency over IPW and should be less susceptibility to bias caused by extreme propensity scores. In the case of a complete outcome, OW has been proved to give the smallest total variance for pairwise comparisons among all balancing weights. The theory and optimality of OW, however, has not been explored with survival outcomes, and will be investigated below.

\section{Theoretical properties}\label{s:theoretical}

We present two main results on the theoretical properties of the proposed weighting estimator \eqref{eq:PSW_Pseudo}. The first result develops a new asymptotic variance expression for the weighted pairwise comparisons of the pseudo-observations, and the second result establishes the efficiency optimality of OW within the family of balancing weights based on the pseudo-observations. 

Below we first outline the main steps of deriving the asymptotic variance. Let $(\Omega,\mathcal{F},\mathcal{P})$ be a probability space and $(\mathbf{D},\Vert\bullet\Vert 
)$ be a Banach space for distribution functions. Specifically, we choose the Banach space to the space of fucnctions of bounded p-variation and the corresponding norm $\Vert\bullet\Vert$ is the p-variation norm. We refer the reader to example 3.2 in \cite{overgaard2017asymptotic} for regularity details. We assume each tuple $\mathcal{O}_{i}=(Z_{i},\bcx_{i},\widetilde{T}_{i},\Delta_{i})$ is an i.i.d draw from the sample space $\mathcal{S}$ in the probability space $(\Omega,\mathcal{F},\mathcal{P})$. Define the Dirac measure $\delta_{(\bullet)}:\mathcal{S}\rightarrow \mathbf{D}$, we write the \emph{empirical distribution function} as $F_{n}=N^{-1}\sum_{i=1}^{N}\delta_{\mathcal{O}_{i}}$ and its limit as $F$. Following \citet{overgaard2017asymptotic}, we use functionals to represent different estimators for the transformed survival outcomes with pseudo-observations. Suppose $\phi_{k}(\bullet;t):\mathbf{D}\rightarrow \mathcal{R}$ is the functional mapping a distribution to a real value, such as the Kaplan-Meier estimator,  $\phi_{1}(F_{N};t)=\widehat{S}(t)$, then each pseudo-observation is represented as $\widehat{\theta}_{i}^{k}(t)=N\phi_{k}(F_{N};t)-(N-1)\phi_{k}(F_{N}^{-i};t)$, where $F_{N}^{-i}$ is the empirical distribution omitting $\mathcal{O}_{i}$. 

To derive the asymptotic variance of estimator \eqref{eq:PSW_Pseudo}, we need to accommodate two sources of uncertainty. The first source stems from the calculation of the pseudo-observations. We consider the functional derivative of $\phi_{k}(\bullet;t)$ at $f\in \mathbf{D}$ along direction $s\in\mathbf{D}$ as $\phi_{k,f}'(s)$, which is a linear and continuous functional, $\{\phi_{k}(f+
s;t)-\phi_{k}(f;t)-\phi'_{k,f}(s;t)\}^{2}=o(||s||_{\mathbf{D}})$. Assuming $\phi_{k}(\bullet;t)$ is differentiable at the true distribution function $F$, we express the first-order influence function of $\mathcal{O}_{i}$ for the pseudo-observation estimator $\hat{\theta}^{k}(t)$ as the first-order derivative along the direction $\delta_{\mathcal{O}_{i}}-F$, denoted by $\phi_{k,i}'(t)\triangleq\phi_{k,F}'(\delta_{\mathcal{O}_{i}}-F;t)$. Similarly, the second-order derivative for the functional $\phi_{k}(\bullet;t)$ at $f$ along direction $(s,w)$ can be defined as $\phi_{k,F}''(s,w;t)$, and the second-order influence function for $(\mathcal{O}_{i},\mathcal{O}_{j})$ is given as $\phi_{k,(l,i)}''(t)\triangleq\phi_{k,F}''(\delta_{\mathcal{O}_{l}}-F,\delta_{\mathcal{O}_{i}}-F;t)$.
To characterize the variability associated with jackknifing, we follow \citet{graw2009pseudo} and \citet{jacobsen2016pseudo_large_sample} to write the second-order von Mises expansion of the pseudo-observations: 
\begin{eqnarray}
\label{eqn:functional_taylor_expansion}
\widehat{\theta}_{i}^{k}(t)=\theta^{k}(t)+\phi_{k,i}'(t)+\frac{1}{N-1}\sum_{l\neq i}\phi_{k,(l,i)}''(t)+R^{k}_{N,i},
\end{eqnarray}
where the first three terms dominate the asymptotic behaviour of $\widehat{\theta}_{i}^{k}(t)$ and the remainder $R^{k}_{N,i}$ vanishes asymptotically because $\lim_{N\rightarrow 0}\sqrt{N}\textup{max}_{i}|R^{k}_{N,i}|=0$ for any $k$. The second source of uncertainty in estimator \eqref{eq:PSW_Pseudo} comes from estimating the unknown propensity scores and hence the weights; such uncertainty is well studied in the causal inference literature and is usually quantified using M-estimation (see, for example, \citet{lunceford2004stratification}). Typically, the unknown propensity score model is parameterized as $e_{j}(\bcx_{i};\bfgamma)$, where the parameter $\bfgamma$ is estimated by maximizing the multinomial likelihood. 
\begin{theorem}
\label{thm:1}
Under suitable regularity conditions specified in Web Appendix A in the supplementary materials, for $k=1,2$, $j,j'\in \mathcal{J}$ and all continuously differentiable tilting function $h(\bcx)$, (a) $\widehat{\tau}_{j,j'}^{k,h}(t)$ is a consistent estimator for $\tau_{j,j'}^{k,h}(t)$; (b) $\sqrt{N}\left\{\widehat{\tau}_{j,j'}^{k,h}(t)-\tau_{j,j'}^{k,h}(t)\right\}$ converges in distribution to a mean-zero normal random variate with variance $\bE\{\Psi_{j}(\mathcal{O}_i;t)-\Psi_{j'}(\mathcal{O}_i;t)\}^{2}/\{\bE(h(\bcx_{i}))\}^2$, where 
\begin{align}\label{eq:if}
\Psi_{j}(\mathcal{O}_i;t)=&\textbf{1}\{Z_i=j\}w_{j}^{h}(\bcx_{i})
\left\{\left(\theta^{k}(t)+\phi'_{k,i}(t)-m_{j}^{k,h}(t)\right)+Q_i\right\}\nonumber\\
&+\bE\left\{\textbf{1}\{Z_i=j\}
\left(\theta^{k}(t)+\phi'_{k,i}(t)-m_{j}^{k,h}(t)\right)\frac{\partial}{\partial \gamma^T}w_{j}^{h}(\bcx_{i})\right\}\bci_{\bfgamma\bfgamma}^{-1}\bcs_{\bfgamma,i},   
\end{align}
$Q_i=(N-1)^{-1}\sum_{l\neq i}\phi_{k,(l,i)}''(t)\textbf{1}\{Z_l=j\}w_{j}^{h}(\bcx_{l})$,  
$\bcs_{\bfgamma,i}$ and $\bci_{\bfgamma\bfgamma}$ are the score function and information matrix of $\bfgamma$, respectively. 
\end{theorem}

Theorem \ref{thm:1} establishes consistency and asymptotic normality of the proposed weighting estimator \eqref{eq:PSW_Pseudo}. In particular, the influence function $\Psi_{j}(\mathcal{O}_i;t)$ delineates two aforementioned sources of variability, with the first and second term characterizing the uncertainty due to estimating the pseudo-observations and the propensity scores, respectively. The jackknife pseudo-observation estimator for $\widehat{\theta}_i^k(t)$ includes information from the rest $N-1$ observations and thus is no longer independent across units. Therefore, derivation of \eqref{eq:if} requires invoking the central limit theorem for U-statistics \citep[cf. Chapter 12 in][]{van2000asymptotic}, and leads to a second-order term, $Q_i$, that properly accommodates the correlation between the estimated pseudo-observations of different units. Theorem \ref{thm:1} immediately suggests the following consistent variance estimator for pairwise comparisons, $\widehat{\bV}\{\widehat{\tau}_{j,j'}^{k,h}(t)\}=\sum_{i=1}^{N}\{\widehat{\Psi}_{j}(\mathcal{O}_i;t)-\widehat{\Psi}_{j'}(\mathcal{O}_i;t)\}/\sum_{i=1}^{N}\widehat{h}(\bcx_{i})^{2}$, where $\widehat{\Psi}_{j}(\mathcal{O}_i;t)$ is defined explicitly in Web Appendix A. In Web Appendix A, we also give explicit derivations of the functional derivatives for each transformation $\nu_k$ when the Kaplan-Meier estimator is used to construct the pseudo-observations as in Section \ref{sec:PO}. This new closed-form  estimator enables fast computation of the variance of the weighting estimator \eqref{eq:PSW_Pseudo} without resampling, a crucial advantage when the sample size is large.  

Several important remarks regarding Theorem \ref{thm:1} are in order.

\begin{Remark}\label{rmk:1}
Without censoring, each pseudo-observation degenerates to the observed outcome, which implies $\widehat{\theta}_i^{k}(t)=\theta^{k}(t)+\phi'_{k,i}(t)=\nu_{k}(T_{i};t)$ and therefore {$Q_i=0$}. In this case, formula \eqref{eq:if} coincides with the influence function derived in \citet{li2019propensity} for complete outcomes.
\end{Remark}

\begin{Remark}\label{rmk:2}
In the presence of censoring, we show in Web Appendix A that ignoring the uncertainty due to estimating pseudo-observations will, somewhat counter-intuitively, \emph{overestimate} the variance of $\widehat{\tau}_{j,j'}^{k,h}(t)$. This insight for weighting estimator is in parallel to \citet{jacobsen2016pseudo_large_sample}, who suggested ignoring the uncertainty due to estimating the pseudo-observations leads to conservative inference for outcome regression coefficients.
\end{Remark}

\begin{Remark}\label{rmk:3}
For $h(\bcx)=1$ (and equivalently the IPW scheme), we show in Web Appendix A that treating the inverse probability weights as known will, also counter-intuitively, \emph{overestimate} the variance for pairwise comparisons; this extends the classic results of \citet{Hirano2003} to multiple treatments. The implications of ignoring the uncertainty in estimating the propensity scores, however, are generally uncertain for other choice of $h(\bcx)$, which can lead to either conservative or anti-conservative inference, as also mentioned in \citet{haneuse2013estimation}. An exception is the randomized controlled trial (RCT), where the propensity score to any treatment group is a constant and thus any tilting function based on the propensity scores reduces to a constant, i.e. $h(\bcx)=\widetilde{h}(e_1(\bcx),\ldots,e_j(\bcx))\propto 1$. In this case, one can still estimate a ``working'' propensity score model and use the subsequent weighting estimator  \eqref{eq:PSW_Pseudo} to adjust for chance imbalance in covariates. Equation \eqref{eq:if} shows that such a covariate adjustment approach in RCT leads to variance reduction for pairwise comparisons, extending the results developed in \citet{Zeng2020} to multiple treatments and censored survival outcomes.
\end{Remark}

\begin{Remark}\label{rmk:4}
Estimator \eqref{eq:PSW_Pseudo} and Theorem \ref{thm:1} can be extended to accommodate covariate dependent censoring: $T_{i}(j)\indep C_{i}(j)|\bcx_{i},Z_i$. In this case, one can consider inverse probability of censoring weighted pseudo-observation  \citep{robins2000correcting_IPC,binder2014pseudo_depedentcensoring}:
\begin{eqnarray}
\label{eq:dependent_censoring_pseudo}
\widehat{\theta}_{i}^{k}(t)=\frac{\nu_{k}(\widetilde{T}_{i};t)\textbf{1}\{C_{i}\geq \widetilde{T}_{i}\wedge t\}}{\widehat{G}(\widetilde{T}_{i}\wedge t|\bcx_{i},Z_{i})},
\end{eqnarray}
where $\widehat{G}(u|\bcx_{i},Z_{i})$ is a consistent estimator of the censoring survival function $G(u|\bcx_{i},Z_{i})=\Pr(C_{i}\geq u|\bcx_{i},Z_{i})$, for example, given by the Cox proportional hazards regression. {There are other possible types of pseudo-observations adjusting for the dependent censoring \citep{binder2014pseudo_depedentcensoring}. We select \eqref{eq:dependent_censoring_pseudo} for to simplify the computation, especially when deriving the consistent variance estimator.} To show the consistency and asymptotic normality of the modified weighting estimator, we can similarly view \eqref{eq:dependent_censoring_pseudo} as a functional mapping from the empirical distribution of data to a real value \citep{overgaard2019dependent_pseudo} and find the corresponding functional derivatives for asymptotic expansion (see Web Appendix A).
\end{Remark}

The following Theorem \ref{thm:2} shows that the overlap weights, similar to the case of non-censored outcomes, lead to the smallest total asymptotic variance for all pairwise comparisons based on the pseudo-observations among the family of balancing weights.
\begin{theorem}\label{thm:2} 
Under regularity conditions in Web Appendix A and assuming generalized homoscedasticity such that $\lim_{N\rightarrow\infty}\bV\{\hat{\theta}_{i}^{k}(t)|Z_{i},\bcx_{i}\}=\bV\{\phi_{k,i}'(t)|Z_{i},\bcx_{i}\} $ is a constant across different levels of $(Z_{i},\bcx_{i})$, the harmonic mean function $h(\bcx)=\left\{\sum_{l\in\mathcal{J}}e_{l}^{-1}(\bcx)\right\}^{-1}$ leads to the smallest total asymptotic variance for pairwise comparisons among all  tilting functions.
\end{theorem}

Theorem \ref{thm:2} generalizes the findings of \citet{crump2006moving}, \citet{li2018balancing} and \citet{li2019propensity} to provide new theoretical justification for the efficiency optimality of the overlap weights, 
$w_{j}^{h}(\bcx)=e_{j}(\bcx) \left\{\sum_{l\in\mathcal{J}}  e_{l}^{-1}(\bcx)\right\}^{-1}$, when applied to censored survival outcomes. Technically this result relies on a generalized homoscedasticity assumption that requires the limiting variance of the estimated pseudo-observations to be constant within the strata defined by $(Z_{i},\bcx_{i})$. This condition includes the usual homoscedasticity for conditional outcome variance as a special case in the absence of censoring. 
Of note, the homoscedasticity condition may not hold in practice, but has been empirically shown to be not crucial for the efficiency property of OW, as exemplified in the simulations by \citet{li2018balancing} and numerous applications. Furthermore, in Section \ref{s:simulations}, we carry out extensive simulations to verify that OW leads to improved efficiency over IPW when generalized homoscedasticity is violated.

We can further augment estimator \eqref{eq:PSW_Pseudo} by an outcome regression model of the pseudo-observations. Specifically, for any time $t$, we can posit treatment-specific outcome models $m_{j}^k(\bcx_{i};\balpha_{j})=\bE\{\widehat{\theta}_{i}^{k}(t)|\bcx_{i},Z_i=j\}$, and define an augmented weighting estimator
\begin{align}
\label{eq:Augmented}
\widehat{\tau}_{j,j',\textup{AUG}}^{k,h}(t)=&
\frac{\sum_{i=1}^{N}\widehat{h}(\bcx_{i})\{m_{j}(\bcx_{i},\widehat{\balpha}_{j})-m_{j'}(\bcx_{i},\widehat{\balpha}_{j'})\}}
{\sum_{i=1}^{N}\widehat{h}(\bcx_{i})}+\nonumber\\
&\frac{\sum_{i=1}^{N}\textbf{1}\{Z_{i}=j\}\{\widehat{\theta}_{i}^{k}(t)-
m_{j}(\bcx_{i},\widehat{\balpha}_{j})\}w_{j}^{h}(\bcx_{i})}
{\sum_{i=1}^{N}\textbf{1}\{Z_{i}=j\}w_{j}^{h}(\bcx_{i})}
-\nonumber\\
&\frac{\sum_{i=1}^{N}\textbf{1}\{Z_{i}=j'\}
\{\widehat{\theta}_{i}^{k}(t)-m_{j'}(\bcx_{i},\widehat{\balpha}_{j'})\}w_{j'}^{h}(\bcx_{i})}
{\sum_{i=1}^{N}\textbf{1}\{Z_{i}=j'\}w_{j'}^{h}(\bcx_{i})},
\end{align}
where $\widehat{\balpha}_{j}$ denotes the estimated regression parameters in the $j$th outcome model. Such an augmented estimator generalizes those developed in \citet{mao2019propensity} to multiple treatments and survival outcomes. 
When $h(\bcx)=1$, i.e. with the IPW scheme, the augmented estimator becomes the doubly-robust estimator for pairwise comparisons. When only $J=2$ treatments are compared, \eqref{eq:Augmented} reduces to the estimator of \citet{wang2018doublerobustsurvival}, and provides an alternative to other doubly-robust estimators studied in, for example, \citet{zhang2012double}. 
For other choices of $h(\bcx)$, the augmented estimator is not necessarily doubly robust, but may be more efficient than weighting alone when the outcome model is correctly specified \citep{mao2019propensity}. 
For specifying an outcome regression model, \citet{andersen2010pseudoSMMR} reviewed a set of generalized linear models appropriate for the pseudo-observations, and discussed residual-based diagnostic tools for checking model adequacy. One can follow their strategies and assume the outcome model as $m_{j}(\bcx_{i};\balpha_{j})=g^{-1}(\bcx_{i}^{T}\balpha_{j})$, where $g$ is a link function. Estimation of $\balpha_{j}$ can proceed with standard algorithms for fitting generalized linear models. For our estimands of interest, we can choose the identity or log link for estimating the ASCE and RACE and the complementary log-log link (resembling a proportional hazards model) for the SPCE \citep{andersen2004regressionpseudo}. Compared to Theorem \ref{thm:1} for the weighting estimator \eqref{eq:PSW_Pseudo}, derivation of the asymptotic variance of \eqref{eq:Augmented} requires considering a third source of uncertainty due to estimating $\balpha_j$ in the outcome model. We sketch the key derivation steps in Web Appendix A. 

\section{Simulation studies}\label{s:simulations}
\emph{\textbf{Simulation design}.}  We conduct simulation studies to evaluate the finite-sample performance of the weighting estimator \eqref{eq:PSW_Pseudo}, and to illustrate the efficiency property of the OW estimator. We generate four pre-treatment covariates: $\bcx_{i}=(X_{1i},X_{2i},X_{3i},X_{4i})^{T}$, where $(X_{1i},X_{2i})^{T}$ are drawn from a mean-zero bivariate normal distribution with equal variance $2$ and correlation $0.25$, 
$X_{3i}\sim \textup{Bern}(0.5)$, and $X_{4i}\sim \textup{Bern}(0.4+0.2X_{3i})$. We consider $J=3$ treatments, with the true propensity score model given by $\textup{log}\{e_{j}(\bcx_{i})/e_{1}(\bcx_{i})\}=\widetilde{\bcx}_{i}^{T}\bfbeta_{j}$, $j=1,2,3$, where $\widetilde{\bcx}_{i} = (1,\bcx_{i}^{T})^{T}$. We set $\bfbeta_{1}=(0,0,0,0,0)^{T}$, $\bfbeta_{2}=0.2\bfbeta_{3}$; two sets of values for $\bfbeta_{3}$ are considered: (i) $\bfbeta_{3}=(-0.4, 0.85, 0.9, 0.45, -0.25)^{T}$ and (ii) $\bfbeta_{3}=(1.2,1.5,1,-1.5,-1)^{T}$, which represent good and poor covariate overlap across groups, respectively. Distribution of the true generalized propensity scores under each specification is presented in Web Figure 1. 



Two outcome models are used to generate potential survival times. Model A is a Weibull proportional hazards model with hazard rate for $T_i(j)$ as $\lambda_j(t|\bcx_{i})=\eta\nu t^{\nu-1}\exp\{L_{i}(j)\}$, and $L_i(j)=\textbf{1}\{Z_{i}=2\}\gamma_{2}+\textbf{1}\{Z_{i}=3\}\gamma_{3}+\bcx_{i}^{T}\bfalpha$. We specify $\eta=0.0001$, $\nu=3$, $\bfalpha=(0,2,1.5,-1,1)^T$, and $\gamma_{2}=\gamma_{3}=1$, implying worse survival experience due to treatments $j=2$ and $j=3$. The potential survival time is drawn using $T_{i}(j)=\left\{\frac{-\log (U_{i})}{\eta \exp(L_{i}(j))}\right\}^{1/\nu}$, where $U_{i}\sim\textup{Unif}(0,1)$. Model B is an accelerated failure time model that violates the proportional hazards assumption. Specifically, $T_{i}(j)$ is drawn from a log-normal distribution $\log\{T_{i}(j)\}\sim \mathcal{N}(\mu,\sigma^2=0.64)$, with $\mu=3.5-\gamma_{2}\textbf{1}\{Z_{i}=2\}-\gamma_{3}\textbf{1}\{Z_{i}=3\}-\bcx_{i}^{T}\bfalpha$. For simplicity, we assume treatment has no causal effect on censoring time such that $C_i(j)=C_i$ for all $j\in\mathcal{J}$. Under completely independent censoring, $C_{i}\sim \textup{Unif}(0,115)$. Under covariate-dependent censoring, $C_{i}$ is generated from a Weibull survival model with hazard rate $\lambda^c(t|\bcx_i)=\eta_c\nu_ct^{\nu_c-1}\exp(\bcx_{i}^{T}\bfalpha_{c})$, where $\bfalpha_{c} = (1,0.5,-0.5,0.5)^{T}$, $\eta_{c}=0.0001$, $\nu_{c}=2.7$. These parameters are specified so that the marginal censoring rate is roughly $50\%$. Neither data generating process assumes generalized homoscedasticity in Theorem \ref{thm:2}, and thus provides an objective evaluation of the efficiency property of OW. 


Under each data generating process, we consider the OW and IPW estimators based on \eqref{eq:PSW_Pseudo}, and focus our comparison here with two standard estimators: the g-formula estimator based on the confounder-adjusted Cox model, and the IPW-Cox model \citep{austin2017performance}. Details of these two and other alternative estimators are included in Web Appendix B of the supplementary materials. While the IPW estimator \eqref{eq:PSW_Pseudo} and the Cox model based estimators focus on the combined population with $h(\bcx)=1$, the OW estimator focuses on the overlap population with the optimal tilting function suggested in Theorem \ref{thm:2}. When comparing treatments $j=2$ (or $j=3$) with $j=1$, the true values of target estimands can be different between OW and the other estimators (albeit very similar under good overlap), and are computed via Monte Carlo integration. Nonetheless, when we compare treatments $j=2$ and $j=3$, the true conditional average effect $\tau_{2,3}^{k}(\bcx;t)=0$ for all $k$, and thus the true estimand $\tau_{2,3}^{k,h}(t)$ has the same value (zero) regardless of $h(\bcx)$. This represents a natural scenario to compare the bias and efficiency between estimators without differences in true values of estimands. We vary the study sample size $N\in\{150,300,450,600, 750\}$, and fix the evaluation point $t=60$ for estimating SPCE ($k=1$) and RACE ($k=2$). We consider $1000$ simulations and calculate the absolute bias, root mean squared error (RMSE) and empirical coverage corresponding to each estimator. To obtain the empirical coverage for OW and IPW, we construct $95\%$ confidence intervals (CIs) based on the consistent variance estimators suggested by Theorem \ref{thm:1}. Bootstrap CIs are used for Cox g-formula and IPW-Cox estimators. Additional simulations comparing OW with alternative regression estimators and the augmented weighting estimators \eqref{eq:Augmented} can be found in Web Appendix C in the supplementary materials.

\begin{figure}[t!]
\centering
\includegraphics[width=0.8\textwidth]{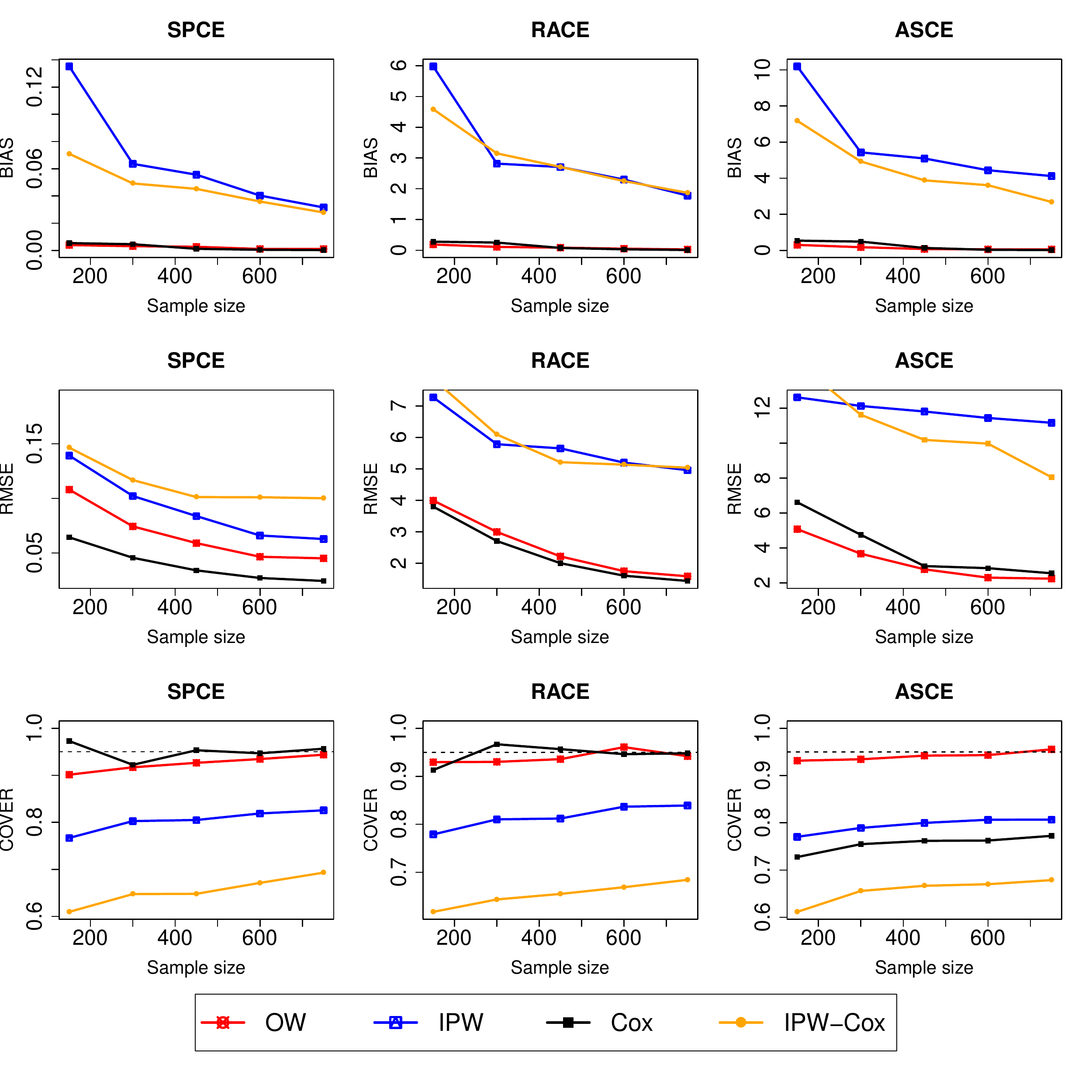}
\caption{\footnotesize Absolute bias, root mean squared error (RMSE) and coverage for comparing treatment $j=2$ versus $j=3$ under poor overlap, when survival outcomes are generated from model A and censoring is completely independent. 
}
\label{fig:outcome_results_poor}
\end{figure}

\emph{\textbf{Simulation results}.} Under good overlap, Web Figure 2 presents the absolute bias, RMSE and coverage for OW, IPW estimators based on \eqref{eq:PSW_Pseudo}, Cox g-formula as well as IPW-Cox estimators, when survival outcomes are generated from model A and censoring is completely independent. Here we focus on comparing treatment $j=2$ versus $j=3$, and thus the true average causal effect among any target population is null. Across all three estimands (SPCE, RACE and ASCE), OW consistently outperforms  IPW with a smaller absolute bias and RMSE, and closer to nominal coverage across all levels of $N$. Due to correctly specified outcome model, the Cox g-formula estimator is, as expected, more efficient than the weighting estimators. However, its empirical coverage is not always close to nominal, especially for estimating ASCE. The IPW-Cox estimator has the largest bias, because the proportional hazards assumption does not hold among any of the target population. Figure \ref{fig:outcome_results_poor} represents the counterpart of Web Figure 2 but under poor overlap. The IPW estimator based on \eqref{eq:PSW_Pseudo} is susceptible to lack of overlap due to extreme inverse probability weights, resulting in extremely large bias, variance and low coverage. The bias and under-coverage remain for IPW even after trimming units with extreme propensities, i.e. with $\max_{j}\{e_{j}(\bcx_{i})\}>0.97$ and $\min_{j}\{e_{j}(\bcx_{i})\}< 0.03$. (Web Figure 3). 
Under poor overlap, OW is more efficient than IPW regardless of trimming, and is almost as efficient as the Cox g-formula estimator for estimating RACE and ASCE. Furthermore, the proposed OW interval estimator carries close to nominal coverage for all estimands. The patterns for comparing treatments $j=2$ and $j=1$ with non-null true average causal effect are similar and presented in Web Figure 7.

Table \ref{tab:cont_simu_results} summarizes the performance metrics for different estimators when the proportional hazards assumption is violated and/or censoring depends on covariates. Similar to Figure \ref{fig:outcome_results_poor}, we focus on comparing treatment $j=2$ versus $j=3$ such that the true average causal effect is null among any target population. When survival outcomes are generated from model B with non-proportional hazards, both the Cox g-formula and IPW-Cox estimators have the largest bias, especially under poor overlap. In those scenarios, OW maintains the largest efficiency, and consistently outperforms IPW in terms of bias and variance. While the coverage of IPW estimator deteriorates under poor overlap, the coverage of OW estimator is robust to lack of overlap. When censoring further depends on covariates, we modify the OW and IPW estimators using \eqref{eq:dependent_censoring_pseudo} where the censoring survival functions are estimated by a Cox model. With the addition of inverse probability of censoring weights, only OW maintains the smallest bias, largest efficiency and closest to nominal coverage under poor overlap across all estimands. Results for comparing treatments $j=2$ and $j=1$ are similar and included in Web Table 1.

\begin{table}[htbp]
\centering
\resizebox{!}{0.55\textheight}{\begin{minipage}{\textwidth}
\caption{\label{tab:cont_simu_results} 
\footnotesize Absolute bias, root mean squared error (RMSE) and coverage for comparing treatment $j=2$ versus $j=3$ under different degrees of overlap. In the ``proportional hazards'' scenario, the survival outcomes are generated from a Cox model (model A), and in the ``non-proportional hazards'' scenario, the survival outcomes are generated from an accelerated failure time model (model B). The sample size is fixed at $N=300$.}
\resizebox{1\textwidth}{!}{
\begin{tabular}{cc@{\hskip 0.3in}cccc@{\hskip 0.3in}cccc@{\hskip 0.3in}cccc}
\hline
&Degree of& \multicolumn{4}{c}{Absolute bias}& \multicolumn{4}{c}{RMSE}&\multicolumn{4}{c}{95\% Coverage}\\
&overlap & OW & IPW & Cox & IPW-Cox & OW & IPW & Cox & IPW-Cox & OW & IPW & Cox & IPW-Cox \\ 
\hline

\multicolumn{14}{c}{Model A, completely random censoring}\\
SPCE & Good & 0.003& 0.006& 0.001& 0.023& 0.062& 0.098& 0.018& 0.091& 0.924& 0.901& 0.949& 0.795 \\ \smallskip 
  & Poor & 0.003& 0.007& 0.005& 0.049& 0.074& 0.102& 0.046& 0.117& 0.917& 0.879& 0.922& 0.647 \\  
 RACE & Good & 0.096& 0.304& 0.086& 1.449& 2.243& 3.379& 1.094& 4.453& 0.937& 0.919& 0.961& 0.797 \\ \smallskip 
  & Poor & 0.109& 0.391& 0.252& 3.151& 2.998& 3.496& 2.709& 6.093& 0.930& 0.901& 0.967& 0.644 \\  
 ASCE & Good & 0.181& 0.354& 0.153& 2.336& 2.916& 4.974& 1.911& 8.959& 0.941& 0.903& 0.849& 0.790 \\ \smallskip 
  & Poor & 0.181& 0.443& 0.490& 4.930& 3.666& 6.373& 4.750& 11.625& 0.934& 0.899& 0.755& 0.656 \\

\multicolumn{14}{c}{Model B, completely random censoring}\\
SPCE & Good & 0.003& 0.005& 0.005& 0.024& 0.087& 0.112& 0.074& 0.176& 0.958& 0.923& 0.749& 0.779 \\ \smallskip 
  & Poor & 0.005& 0.008& 0.016& 0.081& 0.097& 0.118& 0.150& 0.222& 0.941& 0.921& 0.770& 0.712 \\  
 RACE & Good & 0.102& 0.112& 0.239& 1.530& 2.761& 4.304& 4.219& 8.758& 0.960& 0.937& 0.745& 0.787 \\ \smallskip 
  & Poor & 0.105& 0.299& 0.947& 4.646& 3.627& 4.669& 8.653& 11.275& 0.936& 0.929& 0.742& 0.709 \\  
 ASCE & Good & 0.129& 0.443& 0.468& 2.382& 4.238& 7.174& 7.354& 16.583& 0.958& 0.959& 0.846& 0.777 \\ \smallskip 
  & Poor & 0.223& 0.638& 1.661& 7.562& 4.840& 7.189& 15.027& 20.920& 0.961& 0.934& 0.743& 0.705 \\ 
\multicolumn{14}{c}{Model A, covariate dependent censoring}\\
SPCE & Good & 0.002& 0.005& 0.003& 0.038& 0.052& 0.082& 0.047& 0.121& 0.917& 0.889& 0.921& 0.741 \\ \smallskip 
  & Poor & 0.005& 0.007& 0.009& 0.089& 0.060& 0.084& 0.056& 0.149& 0.908& 0.882& 0.881& 0.642 \\  
 RACE & Good & 0.048& 0.154& 0.117& 2.201& 2.773& 3.838& 2.801& 5.382& 0.938& 0.926& 0.908& 0.763 \\ \smallskip 
  & Poor & 0.168& 0.223& 0.532& 4.603& 3.534& 4.207& 3.334& 7.159& 0.935& 0.926& 0.900& 0.634 \\  
 ASCE & Good & 0.055& 0.425& 0.183& 1.161& 5.562& 8.722& 6.005& 36.021& 0.940& 0.909& 0.885& 0.804 \\ \smallskip 
  & Poor & 0.067& 0.568& 1.032& 11.657& 9.557& 9.735& 7.157& 43.651& 0.928& 0.892& 0.752& 0.772 \\  
\multicolumn{14}{c}{Model B, covariate dependent censoring}\\
SPCE & Good & 0.001& 0.001& 0.009& 0.005& 0.050& 0.053& 0.087& 0.075& 0.954& 0.930& 0.699& 0.900 \\ \smallskip 
  & Poor & 0.002& 0.005& 0.012& 0.025& 0.052& 0.082& 0.164& 0.082& 0.925& 0.925& 0.723& 0.896 \\  
 RACE & Good & 0.072& 0.081& 0.498& 0.139& 4.733& 5.879& 4.684& 6.327& 0.954& 0.946& 0.711& 0.850 \\ \smallskip 
  & Poor & 0.109& 0.146& 0.712& 1.594& 6.250& 7.115& 9.092& 7.515& 0.956& 0.955& 0.705& 0.839 \\  
 ASCE & Good & 0.072& 0.258& 0.794& 0.340& 4.436& 5.738& 7.337& 7.756& 0.954& 0.946& 0.835& 0.847 \\ \smallskip 
  & Poor & 0.138& 0.350& 1.339& 1.973& 5.026& 6.503& 13.039& 8.835& 0.955& 0.955& 0.757& 0.847 \\ 
\hline
\end{tabular}
}
\end{minipage}}
\end{table}



We have additionally compared OW with alternative outcome regression estimators similar to \citet{mao2018propensitysurvival}, and the g-formula estimator based on the pseudo-observations. 
These estimators were originally developed with binary treatments, and we adapt them in Web Appendix C to multiple treatments. Compared to the proposed OW estimator \eqref{eq:PSW_Pseudo}, these regression estimators are frequently less efficient and have less than nominal coverage under poor overlap. An exception is the OW regression estimator that generalizes the work of \citet{mao2018propensitysurvival}, which has similar performance to the OW estimator based on \eqref{eq:PSW_Pseudo} when outcome is generated from model A. When outcome is generated from model B, the OW estimator in \citet{mao2018propensitysurvival} is subject to larger bias and RMSE due to incorrect proportional hazards assumption. We have also carried out additional simulations in Web Appendix C to examine the performance of the augmented OW and IPW estimators \eqref{eq:Augmented} relative to the OW and IPW estimators \eqref{eq:PSW_Pseudo}. While including an outcome regression component can notably improve the efficiency of IPW, the efficiency gain for OW estimator due to an additional outcome model is negligible. This speaks to the appeal of the OW estimator because outcome models are almost always misspecified in practice. Additionally, we replicate our simulations under a three-arm RCT similar to \citet{Zeng2020} (see Remark \ref{rmk:3} and Web Appendix C). We confirmed that both OW and IPW estimators are valid for covariate adjustment in RCTs and lead to substantially improved efficiency over the unadjusted comparisons of pseudo-observations in the presence of chance imbalance. Finally, under covariate-dependent censoring, we further compared OW and IPW under a misspecified censoring model and found that OW outperforms IPW across all scenarios. With a misspecified censoring model, OW also maintains nominal coverage except when the failure times are generated from model B and the target estimand is SPCE and ASCE. The details are presented in Web Appendix D.

\section{Application to National Cancer Database}\label{s:data_application}
We illustrate the proposed weighting estimators by comparing three treatment options for prostate cancer in an observational dataset with 44,551 high-risk, localized prostate cancer patients drawn from the National Cancer Database (NCDB). These patients were diagnosed between 2004 and 2013, and either underwent a surgical procedure -- radical prostatectomy (RP), or were treated by one of two therapeutic procedures -- external beam radiotherapy combined with androgen deprivation (EBRT+AD) or external beam radiotherapy plus brachytherapy with or without androgen deprivation (EBRT+brachy$\pm$AD). We focus on time to death since treatment initiation as the primary outcome, and pre-treatment covariates include age, clinical T stage, Charlson-Deyo score, biopsy Gleason score, prostate-specific antigen (PSA), year of diagnosis, insurance status, median income level, education, race, and ethnicity. A total of 2,434 patients died during the study period with their survival outcome observed, while other patients have right-censored outcomes. The median and maximum follow-up time is 21 and 115 months, respectively.

We used a multinomial logistic model to estimate the generalized propensity scores, and visualized the distribution of estimated scores in Web Figure 9. The eleven pre-treatment covariates introduced earlier were considered as confounders that affect both treatment assignment and mortality, and included in the propensity score model. We model age and PSA by natural splines following \citet{ennis2018brachytherapy}, and keep linear terms for all other covariates. We found good overlap across groups regarding the propensity of receiving EBRT+brachy$\pm$AD, but a slight lack of overlap regarding the propensity of receiving RP and EBRT+AD. To assess the adequacy of the propensity score model specification, we checked the weighted covariate balance under IPW and OW based on the maximum pairwise absolute standardized difference (MPASD) criteria, and present the balance statistics in Web Table 4. The MPASD for the $p$th covariate is defined as $\max_{j<j'}\{|\bar{X}_{p,j}-\bar{X}_{p,j'}|/S_{p}\}$, where $\bar{X}_{p,j}=
\sum_{i=1}^{N}\textbf{1}\{Z_{i}=j\}X_{i,p}w^h_{j}(\bcx_{i})/\sum_{i=1}^{N}\textbf{1}\{Z_{i}=j\} w^h_{j}(\bcx_{i})$ is the weighted covariate mean in group $j$, and $S_{p}^{2}=J^{-1}\sum_{j=1}^{J}S_{p,j}^{2}$ is the unweighted sample variance averaged across all groups. Both IPW and OW improved covariate balance compared to no weighting. Of note, while OW with logistic propensity scores leads to exact covariate balance for $J=2$ groups \citep{li2018balancing}, OW with multinomial logistic propensity scores do not guarantee exact covariate balance among $J\geq 3$ groups \citep{li2019propensity}. Nonetheless, Web Table 4 shows that OW still leads to consistently smaller MPASD compared to IPW, with values below the usual 0.1 threshold across all covariates.

\begin{figure}[t!]
\begin{subfigure}[b]{1\textwidth}
\centering
\includegraphics[scale = 0.45]{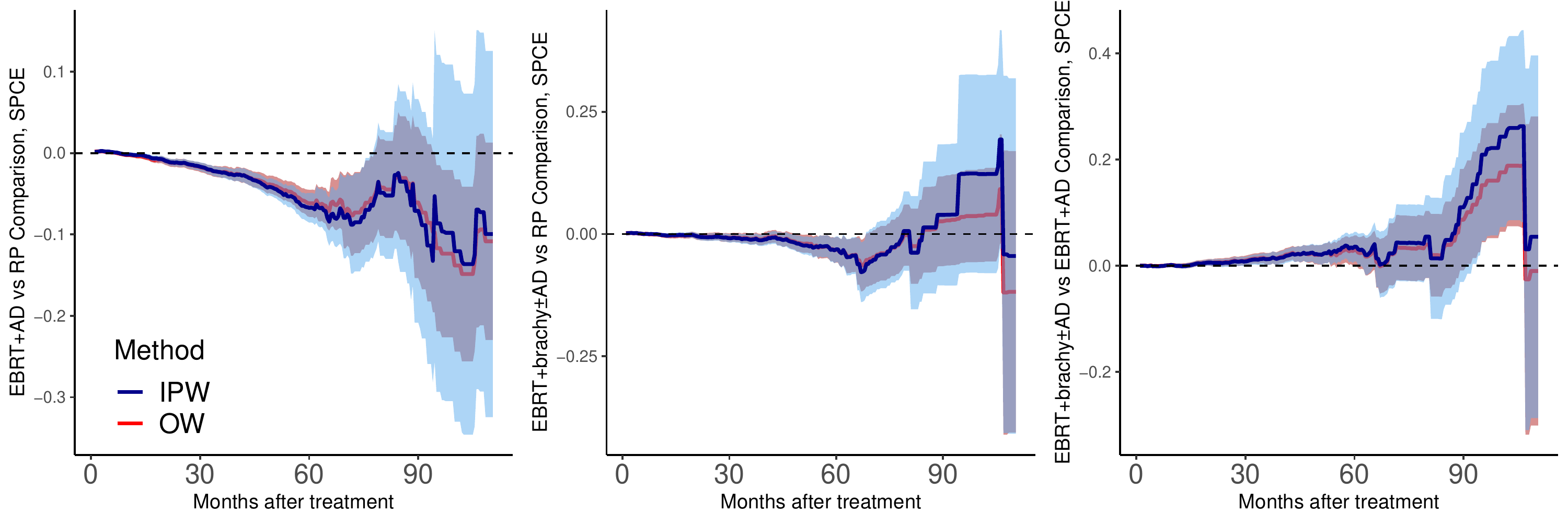}
\caption{Estimated SPCE as a function of time $t$ across three treatment groups.}
\label{fig:spce_diff_application}
\end{subfigure}
\begin{subfigure}[b]{1\textwidth}
\centering
\includegraphics[scale=0.45]{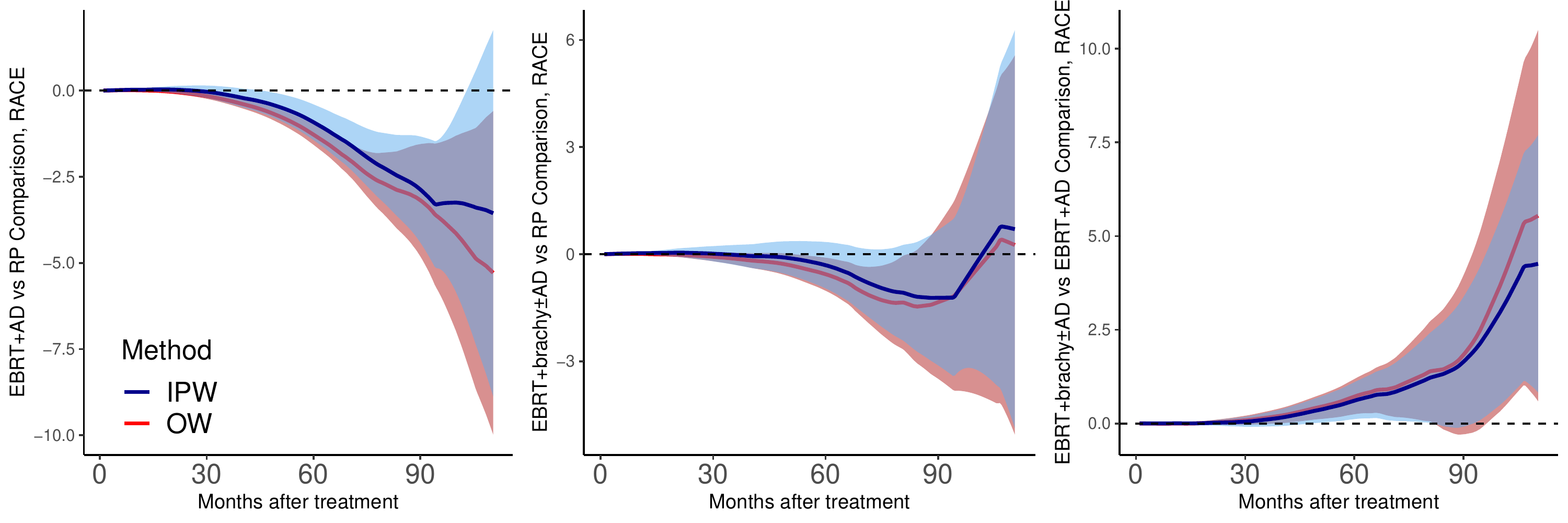}
\caption{Estimated RACE as a function of time $t$ across three treatment groups.}
\label{fig:race_diff_application}
\end{subfigure}
\caption{\footnotesize Point estimates and 95\% confidence bands of SPCE and RACE as a function of time from the pseudo-observations-based IPW and OW estimator in the prostate cancer application in Section \ref{s:data_application}.}
\end{figure}

Web Figure 10 presents the estimated causal survival curves for each treatment, $\bE\{h(\bcx)\textbf{1}\{T_{i}(j)\geq t\}\}/\bE(h(\bcx))$, along with the 95\% confidence bands in the combined population (corresponding to IPW) and the overlap population (corresponding to OW). We chose 220 grid points equally spaced by half a month for this evaluation. The estimated causal survival curves among the two target populations are generally similar, which is expected given there is only a slight lack of overlap. The surgical treatment, RP, shows the largest survival benefit, followed by the radiotherapeutic treatment, EBRT+brachy$\pm$AD, while EBRT+AD results in the worst survival outcomes during the first 80 months or so. Importantly, the estimated causal survival curves for the RP and EBRT+brachy$\pm$AD crossed after month 80, suggesting potential violations to the proportional hazards assumption commonly assumed in survival analysis. Figure \ref{fig:spce_diff_application} and \ref{fig:race_diff_application} further characterized the the SPCE and RACE as a function of time $t$ with the associated 95\% confidence bands. Evidently, the SPCE results confirmed the largest causal survival benefit due to RP, followed by EBRT+brachy$\pm$AD. The associated confidence band of SPCE from OW is narrower than that from IPW and frequently excludes zero. While the analysis of the pairwise RACE yielded similar findings, the efficiency of OW over IPW became more relevant when comparing RP and EBRT+brachy$\pm$AD. Specifically, the confidence band of RACE from OW excludes zero until month 80, while the confidence band of RACE from IPW straddles zero across the entire follow-up period. This analysis shed new light on the significant causal survival benefit of RP over EBRT+brachy$\pm$AD at the 0.05 level in terms of the restricted mean survival time, which was not identified in previous analysis.

In Web Table 4, we also reported the SPCE and RACE using the IPW and OW estimators, as well as the Cox g-formula and IPW-Cox estimators at $t=60$ months, i.e. the 80th quantile of the follow-up time. All methods conclude that RP leads to significantly lower mortality rate at 60 months than EBRT+AD. Compared to IPW, OW provides similar point estimates and no larger variance estimates. Consistently with Figure \ref{fig:race_diff_application}, the smaller variance estimate due to OW (compared to IPW) leads to a change in conclusion when comparing EBRT+brachy$\pm$AD versus RP in terms of RACE at the 0.05 level and confirms the significant treatment benefit of RP. The Cox g-formula and IPW-Cox estimators sometimes provide considerably different results than weighting estimators based on \eqref{eq:PSW_Pseudo}, as they assumed proportional hazards which may not hold (the estimated causal survival curves crossed in Web Figure 10). Overall, we found that, compared to RP, the two radiotherapeutic treatments led to a shorter restricted mean survival time ($1.2$ months shorter with EBRT+AD and $0.5$ month shorter with EBRT+brachy$\pm$AD) up to five years after treatment. The 5-year survival probability is also 6.7\% lower under EBRT+AD and 3.1\% lower under EBRT+brachy$\pm$AD compared to RP. 

\section{Discussion}\label{s:discussion}
We proposed a class of propensity score weighting estimators for survival outcomes based on the pseudo-observations. These estimators are applicable to several different target populations, survival causal estimands, as well as binary and multiple treatments. We also extended our estimators to accommodate covariate dependent censoring and augmentation with outcome models. 
Previous studies rely on bootstrap for variance estimation of similar weighting estimators, which is computationally intensive when combined with the jackknife pseudo-observations. We establish the asymptotic properties of our estimators to motivate a new closed-form variance estimator that takes into account of the uncertainty due to both pseudo-observations calculation and propensity score estimation; this allows valid and fast inference in large observational data. Within the family of balancing weights, we further established the optimal efficiency property of the overlap weights, expanding the theory of overlap weights to survival outcomes. 

An important step in propensity score analysis is to specify the propensity score model. Because the goal of weighting is to balance confounders and removes bias, the weighted covariate balance is routinely used to check whether a propensity score model is adequately specified, and an iterative checking-fitting procedure has been conventionally used to improve model specification. Of note, with $J=2$ treatments, the overlap weights obtained from logistic propensity score model reduce the absolute standardized difference for each covariate to zero, which represents a challenge in operationalizing the iterative checking-fitting procedure \citep{mao2019propensity}. As a potential remedy, one may consider alternative balance metrics such as the weighted differences in empirical distribution function, as in \citet{mccaffrey2013tutorial}. With $J\geq 3$ treatments, the overlap weights generally do not reduce the MPASD balance metric to zero, which suggests that the iterative checking-fitting procedure based on weighted mean balance remains feasible to improve the generalized propensity score model fit. However, because overlap weights often result in relative satisfactorily balance among treatment groups compared to IPW for almost any specification of the generalized propensity score model, a tailored rule of thumb for adequate weighted balance would be of interest and remains an important topic for future research.

The proposed weighting estimators can be extended in several directions. First, while we have focused on estimands on the difference scale, it is straightforward to adapt our weighting estimators to accommodate ratio estimands, which are also of interest in practice. For example, we can write $m_j^{k,h}(t)=\frac{\int_{\mathcal{X}}m_{j}^{k}(\bcx;t)f(\bcx)h(\bcx)\mu(d\bcx)}{\int_{\mathcal{X}}f(\bcx)h(\bcx)\mu(d\bcx)}$, and define the pairwise ratio estimands as $\delta_{j,j'}^{k,h}(t)=m_j^{k,h}(t)/m_{j'}^{k,h}(t)$, $\forall~j\neq j'$. Point identification of $\delta_{j,j'}^{k,h}(t)$ therefore boils down to estimating the average potential outcomes $m_j^{k,h}(t)$ for each $j$ using pseudo-observations, and variance calculation can proceed by applying Delta method to Theorem \ref{thm:1}. In addition, one may further exploit the relationship between survival function and hazard function to define causal hazard difference by $-d\tau_{j,j'}^{k=1,h}(t)/dt$. Inference with this type of estimands, however, requires additional research because our estimator for $\tau_{j,j'}^{k=1,h}(t)$ is non-smooth in $t$. Second, under the covariate dependent censoring, our proposed estimator requires computing pseudo-observations under inverse probability of censoring weighting (IPCW) as in Remark \ref{rmk:4}, which may be inefficient just as using IPW for balancing weights. When the inverse probability of censoring weights are estimated by the Cox model, improvement is possible, for example, by smoothing the baseline hazard estimator to provide potentially more efficient estimation of $\widehat{G}(\widetilde{T}_{i}\wedge t|\bcx_{i},Z_{i})$ and hence the weights \citep{anderson1980smooth}. Alternatively, it may be interesting to develop an augmented-IPCW (hence doubly-robust) pseudo-observation estimator along the lines of doubly-robust censoring unbiased transformation \citep{rubin2007doubly}, which tends to be more efficient than IPCW alone. Adapting these techniques for constructing pseudo-observations is beyond the scope of this work and requires additional research. Finally, \citet{wallace2015doubly} studied OW in constructing the optimal dynamic treatment regimen (DTR) under an additive structural mean model, and demonstrated the efficiency gain over IPW via simulations. Their approach has recently been extended to an additive structural survival model \citep{simoneau2020estimating}. We conjecture that the pseudo-observation approach combined with OW can be a useful alternative to \citet{simoneau2020estimating} in identifying survival DTR under the dynamic weighted ordinary least squares framework.




\section*{Supplementary Materials}

The online supplementary materials include the Web Appendix A-F with technical details and additional simulations, as well as Web Tables and Figures referenced in Section \ref{s:simulations} and \ref{s:data_application}. We provide reproducible R code 
and Web Appendix at \url{https://github.com/zengshx777/OW_Survival_CodeBase}.

\section*{Acknowledgement}
The authors thank the Editor, Associate Editor, and two anonymous referees for constructive suggestions, which greatly improve the exposition of this work.

\par



\bibhang=1.7pc
\bibsep=2pt
\fontsize{9}{14pt plus.8pt minus .6pt}\selectfont
\renewcommand\bibname{\large \bf References}
\expandafter\ifx\csname
natexlab\endcsname\relax\def\natexlab#1{#1}\fi
\expandafter\ifx\csname url\endcsname\relax
  \def\url#1{\texttt{#1}}\fi
\expandafter\ifx\csname urlprefix\endcsname\relax\def\urlprefix{URL}\fi

 \bibliographystyle{chicago}      
 \bibliography{OW_Survival}   





\end{document}